\documentclass{article}
\usepackage{graphicx}
\begin{document}
\title{Quantum communication, reference frames
and gauge theory}
\author{S.J. van Enk\\
Bell Labs, Lucent Technologies\\
600-700 Mountain Ave,
Murray Hill, NJ 07974}
\date{\today}
\maketitle
\begin{abstract}
We consider quantum communication
in the case that the communicating parties not only do not share a reference frame but
use imperfect quantum communication channels, in that each channel
applies some fixed but unknown unitary rotation to each qubit. 
We discuss similarities and differences between reference frames within
that quantum communication model and gauge fields
in gauge theory. We generalize the concept of {\em refbits} and 
analyze various quantum communication protocols within the communication model.
\end{abstract}
\section{Introduction}
In recent years a lot of attention has been paid to the role reference frames
play in quantum communication \cite{direction}--\cite{restric}. 
We may distinguish four different types of studies.
First, many papers consider the resources needed to {\em establish} a shared reference frame \cite{direction}.
For example, a spatial reference frame between Alice and Bob can be established by transmitting spin-1/2 particles,
their spin encoding information about direction. 
The questions that have been considered in that case are whether transmitting entangled 
qubits (here a spin-1/2 particle is viewed as a qubit)
is beneficial and how the fidelity (i.c. the
overlap between Alice's and Bob's private reference
frames) approaches unity with the number of qubits sent.
Similarly one can consider synchronizing clocks 
by transmitting ``ticking'' qubits (superpositions of two quantum states
with different energy) \cite{tick} as establishing a different type of reference frame.

Second, there is a connection between superselection rules and reference frames. More precisely,
the lack of a reference frame leads to an effective superselection rule and, conversely,
a superselection rule can be effectively lifted by setting up a shared reference frame.
All this was noted long ago in \cite{aha} but in recent years the question arose as to whether
superselection rules can be exploited in cryptographic protocols. After all, if an eavesdropper in a cryptographic protocol is not able to perform certain desired operations as a result of superselection rules,
it may enhance the security of that protocol. Unfortunately, 
the answer is negative, precisely because a superselection rule can be effectively lifted.
The proof of that statement is more involved than might be apparent and can be found in \cite{super}.

Third, there are new types of resources associated with reference frames, or the lack thereof.
Although in many cases, for instance in teleportation protocols, it is tacitly assumed an explicit isomorphism 
has been established
between the Hilbert spaces for qubits used by different observers, that is in fact not always
trivial. Hence the ability to establish an isomorphism between different Hilbert spaces is a useful resource 
that until recently stayed hidden \cite{jmod,lasers}.
In order to quantify that resource, Ref.~\cite{supers} introduced
the local
variance in particle number, 
which acts as a resource when a particle-number superselection rule holds. 
Alternatively this quantity can be translated into a {\em refbit} \cite{refbit}, a unit of sharing a reference frame,
in the case that observers agree on the definition of the states $|0\rangle$ and $|1\rangle$ but not on the relative phase
between those two states.
Thus one can quantify the partial presence of a shared reference frame. 
In addition, a {\em privately} shared reference frame between two observers is a useful resource
that can be exploited for
secret communication \cite{private}.

Fourth, there is the question whether the definition of entanglement has to be modified when restrictions exist
on the allowed operations \cite{restric},
such as in the presence of
superselection rules \cite{supers,vaccaro} or in the absence of a shared reference frame \cite{0110}. After all, entanglement
is only a useful concept if it can be used for things like violating Bell inequalities or
teleportation. Both teleportation and Bell-inequality violations
require certain measurements. If those measurements are impossible to perform because of a superselection
rule it makes perfect sense to modify the definition of entanglement. If those measurements
are impossible to perform without a shared reference frame, then one still can 
(partially) convert ``useless'' entanglement into useful entanglement by establishing (partially)
a reference frame by using the above-mentioned {\em refbits}.
 
In all the above papers
the assumption is made that the communicating
parties, although not sharing a reference frame, do share a perfect
quantum communication channel. However, it is hard to see how the communicating parties
can be sure to have a perfect communication channel when they do not
share a reference frame. And so we will instead assume
that some fixed unitary rotation is applied to the qubit upon traversing the communication 
channel. This unitary rotation is assumed to be measurable, at least up to redefinitions
of local reference frames.

In spite of the fact that we do not even consider decoherence in the channel (i.e., we assume
no entanglement
is created between the transmitted qubit and some other system)
this modification is nontrivial. In particular, allowing an asymmetry between 
quantum communication from Alice to Bob and communication from Bob to Alice
leads to geometrical considerations analogous to those encountered 
in gauge theories (see Section 4). For example, gauge transformations can be directly translated
into changes of reference frames.

This paper is organized as follows. In Section 2 we define our communication model
where multiple parties share classical and quantum channels and possess their own private reference frames.
All channels and reference frames are assumed stationary in time.
In Section 3 we define three different types of observables naturally arising in the context of
the communication model. The three different types are defined by the answers to two questions:
is the observable measurable by only one observer or by multiple observers, and is the observable independent of reference frames?

In Section 4 we discuss analogies and differences between (Yang-Mills) gauge theory
and the quantum communication model of Section 2. We also discuss gauge-invariant measurements, not related to
the observables discussed in Section 3. 

In Section 5 we discuss three different resources
for quantum communication protocols. One resource generalizes the {\em refbit} to the present communication model,
the other two resources
are 2-qubit and 3-qubit entangled states. One question considered for the latter resources is when they are equivalent, i.e., when the observers can convert one type of state into another by local operations and classical
communication (LOCC).
Since we assume here the observers do not share reference frames, not all such states are equivalent, even when they would be 
equivalent in the case of shared reference frames. Moreover, we find that
2-qubit entangled states behave very differently in that respect
than do 3-qubit entangled states.

Finally, in Section 6 we study the
modifications of standard quantum communication protocols such as quantum data hiding, superdense coding, and quantum bit commitment. The latter is still not possible--this follows from \cite{super}-- 
but it is perhaps illuminating to see once more why bit commitment fails in a concrete example.
In addition, the example from Section 6.3 is a bit different than standard bit commitment attempts.
Section 7 concludes and summarizes.
\section{Communication model}
\subsection{Quantum communication channels}
We assume we have multiple observers with the standard names Alice, Bob, Charlie etc., with sub- and superscripts pertaining to those
observers abbreviated to
$A,B,C,\ldots$.
The observers communicate over two-way
classical channels and {\em separate} one-way quantum channels. This is 
unlike in Refs.~\cite{encoding,refbit,harrow} where the quantum
communication channel is being used for classical communication as well.
This is meant to imply we do not consider sending a classical bit of information to be 
a difficult task in the present context.

Each pair of observers $k,l=A,B,C,\ldots$ 
thus shares {\em two} quantum communication channels
characterized by two unitaries $V_{kl}\in {\rm SU}(2)$
and $V_{lk}\in {\rm SU}(2)$, 
describing
the transformation of a qubit state when it is
sent from $l$ to $k$ or {\em vice versa}, respectively. 
In general we have {\em neither} $V_{kl}= V_{lk}$
{\em nor} $V_{kl}= V^\dagger_{lk}$.  The latter relation could
hold if the ``same'' physical channel is used
to communicate in both directions. For instance, in the ``plug-and-play''
system developed for quantum key distribution \cite{plug} an essential role is played
by Faraday rotators that ensure a photon that has traveled from Bob to Alice will reverse
its (unknown) polarization rotation when traveling back to Bob over the same channel.
In such a case, when one does have the relation $V_{AB}= V^\dagger_{BA}$, one could 
transform away the action of the channel
and incorporate it into the reference frames of Alice and Bob. But here we do {\em not} make that assumption.
\subsection{Reference frames}
Each observer possesses his/her own local reference
frame, which defines the local orientation of the Bloch sphere, i.e.
what the observer means by ``$|0\rangle$'' and ``$(|0\rangle+|1\rangle)/\sqrt{2}$'', etc.,
and which similarly determines what is meant by ``bit-flip'' and ``phase-flip'' operations.
We assume observers have no knowledge of the other observers' reference frames.

More precisely, then, each observer $k=A,B,C\ldots$ possesses a local reference frame that defines a 
local basis $\{|0_k\rangle,|1_k\rangle\}$.
Of course, only {\em relative} orientations of reference frames
are observable. We denote the basis
transformations that specify
the relative orientations of the frames of reference
of observers $k$ and $l$ by $R_{kl}\in{\rm SU}(2)$. 
We thus explicitly factor out the unobservable ${\rm U}(1)$ transformations 
from the full group of unitaries ${\rm U}(2)$. 
These unitary operators satisfy by definition
\begin{equation}\label{defR}
\langle a_k|R_{kl}|b_l\rangle=\delta_{ab}\,\,\,{\rm for}\,a,b=0,1.
\end{equation}
We have the relations
\begin{eqnarray}
R_{kl}R_{lm}&=&R_{km},\nonumber\\
R_{kl}^\dagger&=&R_{lk}.
\end{eqnarray}
Given (\ref{defR}) we may use the notation
\begin{equation}
|\psi_k\rangle=R_{kl}|\psi_l\rangle,
\end{equation}
to connect the descriptions of different observers
for any state $|\psi\rangle$.
Similarly, we can write the relation between
unitary operations $U$
as applied by different observers:
\begin{equation}
U_{kk}=R_{kl}U_{ll}R_{lk}.
\end{equation}
Depending on the physical implementation of qubits used, one may impose restrictions on the form of the operators $R_{kl}$.
For instance, when using polarized photons
for quantum communication the right-hand and left-hand circular polarization states
can be locally defined with respect to the propagation direction of the light 
since Nature does distinguish between ``left'' and ``right''. (Here we assume we can ignore
the possibility that
the spacetime structure of the universe is that of an unorientable manifold.)
In this case, the observers could agree on choosing left-hand and right-hand circular polarization
to encode $|0\rangle$ and $|1\rangle$ 
without having to share a reference frame. In that case
the form of $R_{kl}$ would be restricted to rotations around the polar axis ($z$ axis), i.e.,
\begin{equation}\label{z}
R_{kl}=\exp(i\phi_{kl}\sigma_z/2),	
\end{equation}
with $\sigma_z$ the generator of rotations around the $z$ axis and $\phi_{kl}$
the phase difference characterizing the relative rotation between the two reference frames of $k$ and $l$.
Here, however, we will {\em not} make the assumption (\ref{z}) on reference frames.
Note also that propagation of polarized photons through a standard optical fiber
can certainly transform 
left-hand into right-hand polarization, and hence no
restriction would be placed on the unitarires $V_{kl}$.
\subsection{Time evolution}
We assume that all the unitaries introduced so far
are stationary. That is, 
both quantum channels and reference frames are assumed stationary.
This description, perhaps despite appearances, does
include the case where ``reference frames'' are clocks.
What is stationary in that case is, e.g.,
the frequency of the clock used and what is unknown is the time-offset between different clocks.

Similarly, there may be a nontrivial time evolution for traveling qubits,
but that evolution is assumed to be stationary, too.
That time evolution can then in fact be easily transformed away.
In most cases this leads to a simpler description. On the other hand,
retaining the time evolution allows one to use coordinate-independent expressions.
Typically, we will transform away the time evolution of a traveling qubit, but give covariant expressions
when appropriate.
\section{Observables}
The unitaries $R_{kl}$ and $V_{kl}$ are not
observable by themselves. In fact, one may even get the impression that
$V_{kl}$ can somehow be redefined to be the identity by absorbing its effects
in the definition of $R_{kl}$. If that were the case, there would be no point to this paper.
That this 
is not so we will see in this Section \footnote{This is not meant to imply the present paper is not pointless,
but only that one cannot set the unitaries $V_{kl}$ equal to the identity.}.
\subsection{Private and public observables}
We will distinguish three types of observables: (i) those 
dependent on the reference frame(s) of other observer(s),
(ii) those independent
of other observers' reference frames, but dependent on the private 
reference frame of the observer, 
and (iii) those 
that are independent of any reference frames.

In the context of communication protocols 
a reasonable working assumption is that all observers have learned everything there is to learn
about the reference-frame-independent observables of type (iii). 
In particular, we will assume that Alice knows reference-frame-independent
observables involving the channels between Bob and Charlie. 

On the other hand, we will assume that the observers know nothing 
about quantities that depend on other observers' reference frames.
So, the latter quantities, of both types (i) and (ii),
can be considered ``private'' variables, while the type (iii) observables are ``public.''
So the three types of observables may be called
private reference-frame dependent (of type (i)),
private reference-frame independent (of type (ii)), and
public reference-frame independent (of type (iii)).

The assumption that all public observables are known to all observers
identifies another resource that is being used
in standard quantum communication protocols. Indeed, such a resource is used implicitly when one assumes
that all quantum communication channels are perfect.
Here we do not explore the issue of quantifying that resource further. This would involve
assuming that observers do {\em not} know anything
about the public reference-frame-independent observables, assessing how
that lack of knowledge affects standard communication protocols such as teleportation,
superdense coding and quantum data hiding, and quantifying how much knowledge is
needed to overcome obstacles in the way of teleportation etc.
\subsection{Private reference-frame-dependent observables} 
The first type of observables allows one observer
to learn something about the reference frame chosen
by another observer, relative to his/her own. 
For example, if Alice prepares a state $|\psi_A\rangle$ and
sends it to Bob, then Bob can measure matrix elements
of the form
\begin{equation}\label{g1}
\langle \phi_B|V_{BA}R_{AB}|\psi_B\rangle.
\end{equation}
This procedure must make use of a classical communication (either from Alice to Bob
or {\em vice versa})
containing a classical description of the state that Alice is supposed to send.
In principle this can be done by transmitting two complex coefficients
$(a_0,a_1)$ describing a qubit state $a_0|0_A\rangle+a_1|1_A\rangle$. In
practice though, it is sufficient to choose from only a small set of states,
such as $\{|0\rangle, |1\rangle, (|0\rangle\pm|1\rangle)/\sqrt{2},
(|0\rangle\pm i|1\rangle)/\sqrt{2}\}$, as long as the set is sufficient
for doing full tomography on a qubit. Thus in order to determine matrix elements
(\ref{g1}) of the unitary $V_{BA}R_{AB}$ 
only a few bits of classical information
need to be sent at a time.

We can also assume Bob applies some unitary $U$, not equal to the identity,
and then returns the qubit to Alice. This leads to observables of the form
\begin{equation}\label{g2}
\langle \phi_A| V_{AB}U_{BB}V_{BA}|\psi_A\rangle,
\end{equation}
which can be measured by Alice.
This requires classical communication of the matrix elements
$\langle a_B|U_{BB}|b_B\rangle$ of the unitary that Bob applies. 
Here too, the unitaries may be chosen from a small set of gates, such as the Pauli gates $X,Y,Z$
(which are up to an overall phase shift equal to the Pauli matrices $\sigma_j$ for $j=x,y,z$, except now seen as unitaries rather
than as generators of unitaries). We conclude that only a small number of bits of communication is required
for observers to measure private reference-frame-dependent observables.
\subsection{Private reference-frame-independent observables}
The second type of observables is
independent of the choices of other observers' reference frames,
and always involves qubits that travel on a closed loop starting and finishing at one and the same observer.
The simplest kind is, e.g., where Alice sends a qubit
to Bob who returns it to Alice without messing with its state. This allows Alice to measure
the matrix elements
\begin{equation}\label{holo}
\langle \phi_A|V_{AB}V_{BA}|\psi_A\rangle.
\end{equation}
This observable shows the definition of $V_{AB}$ cannot be absorbed into the definition
of $R_{AB}$ and $R_{BA}$, it being independent of Bob's reference frame.
In particular, $V_{AB}$ and $V_{BA}$ cannot be set equal to the identity.
The observable does depend on Alice's reference frame. That is,
although Alice could communicate classically the entries of the unitary matrix
(\ref{holo}) to other observers, they cannot reconstruct the same unitary operator from that description. 

Similarly, Alice can measure matrix elements like
\begin{equation}\label{holo2}
\langle \phi_A|V_{AC}V_{CB}V_{BA}|\psi_A\rangle,
\end{equation}
where the qubit commutes from Alice to Bob to Charlie before returning to Alice (and so on for 
multiple stops at various observers).
\subsection{Public reference-frame independent observables}
From the two observables given in the preceding subsection it is easy to construct
quantities that are independent of Alice's reference frame. Namely Alice can measure the
trace of those observables:
\begin{equation}\label{holok}
\sum_{k=0,1}\langle k_A|V_{AB}V_{BA}|k_A\rangle.
\end{equation}
and
\begin{equation}\label{holo2k}
\sum_{k=0,1}\langle k_A|V_{AC}V_{CB}V_{BA}|k_A\rangle.
\end{equation}
But those quantities can in fact also be measured by Bob and Charlie, thanks
to the cyclic property of the trace. Hence the name public observables. 
\section{Quantum communication as a gauge theory}
\subsection{Holonomies and Wilson loops}
The quantity appearing in Eq.~(\ref{holo})
can be written as a holonomy of a connection, i.e.
a path-ordered exponential of a gauge field $\vec{A}$ over 
the path of the qubit in space
\begin{eqnarray}\label{space}
V_{AB}V_{BA}={\cal P}\exp \left( 
i\oint_{\tilde{C}} q\vec{A}\cdot{\rm d}\vec{r}, 
\right)
\end{eqnarray}
where the integral is taken over the closed circuit $\tilde{C}$ in space the qubit
traverses from Alice to Bob and back. Here $q$ is the ``charge''
of the qubit, and $\vec{A}$ takes its value in the Lie algebra su(2).  
The path-ordering operator ${\cal P}$ is necessary to account for the fact that 
fields $\vec{A}$ at different positions
do not necessarily commute.

In writing down a path-integral in ordinary space we explicitly transformed away
the time evolution of the qubit from (\ref{space}).
By reintroducing that time evolution we can rewrite
(\ref{space}) in a covariant way:
First we write
\begin{eqnarray}\label{connection}
V_{BA}={\cal P}\exp \left( 
i\int_{x_A}^{x_B}qA_\mu\,{\rm d}x^\mu, 
\right)
\end{eqnarray}
in terms of a gauge field $A_\mu=(A_t,\vec{A})$, that includes a time-component now. 
Second we explicitly
include the time evolution
along an observer's world line to write a closed spacetime integral
\begin{eqnarray}\label{connectiono}
V_{AB}V_{BA}={\cal P}\exp \left( 
i\oint_C qA_\mu\,{\rm d}x^\mu, 
\right)
\end{eqnarray}
where the integral is over the closed contour $C$ along which the qubit
travels from Alice to Bob and back, and backwards in time in Alice's
frame. 

The observable Eq.~(\ref{connectiono}) still depends on what the starting point
of the traveling qubit is on the closed contour. That is, only Alice can measure
that quantity, and the similar quantity defined for Bob is in general different,
$V_{AB}V_{BA}\neq V_{BA}V_{AB}$ (and
that's why in the previous Section we called this a {\em private} observable). 
If we define
\begin{eqnarray}\label{wilson}
W\equiv{\rm tr} (V_{AB}V_{BA})={\rm tr}{\cal P}\exp \left( 
i\oint_C qA_\mu\,{\rm d}x^\mu, 
\right)
\end{eqnarray}
the resulting quantity becomes independent of the starting point, because of the
cyclic property of the trace. This is in fact what one calls a Wilson loop. It is measurable
both by Alice and Bob, and it is a {\em public} reference-frame independent observable,
in the language of the preceding Section.
\subsection{Matter and field degrees of freedom}
In the pretty much standard case where photon polarization is used to implement a qubit for communication,
there is a reversal of roles of field and matter degrees of freedom
as compared to usual gauge theories. This reversal manifests itself in several different ways:

In gauge theories $A_\mu$ always describes a field
and that field acts on the quantum state of a material particle, such as a proton. But
in our present model the qubit describes a field degree of freedom and $A_\mu$ describes
how matter (e.g. an optical fiber) acts on the polarization degree of a photon. 

In addition,
in our present context the photon is the quantum object
while the matter degrees of freedom, as they are treated here, are classical. In gauge theories
it is the matter particle that's always treated quantum-mechanically while the field may be described classically.
Of course, the main aim of Yang-Mills theory is to quantize that field, too.
Similarly, we could attempt to quantize the gauge field $A_\mu$
but we leave that can of worms closed.

Finally, whereas in gauge theories the material particle is charged, and the gauge field
interacts with the particle through the charge, here the qubit is ``charged'', even if it
is implemented by a photon.
\subsection{Reference frame changes as gauge transformations}
Wilson loops are invariant under infinitesmal gauge transformations of the form
(using the covariant form now)
\begin{equation}\label{gauge}
\delta A^a_\mu=\frac{1}{q}\partial_\mu \alpha^a+f^{abc}A^b_\mu\alpha^c,
\end{equation}
for arbitrary (infinitesimal) $\alpha^a(x_\mu)$ where we wrote
\begin{equation}
A_\mu=A^a_\mu t_a,
\end{equation}
with $t_a=\sigma_a/2$ the generators of ${\rm SU}(2)$ and 
the structure constants are given by $f^{abc}=\epsilon^{abc}$
in terms of the anti-symmetric Levi-Civita symbol.
Equivalently, the gauge transformation can be written as
\begin{equation}\label{Ag}
A_\mu(x)\mapsto U(x)\left( A_\mu+i\frac{1}{q}\partial_\mu  \right)U^\dagger(x),
\end{equation}
with $U(x)$ any unitary, varying smoothly with spacetime. 
But as we noted before, the same quantity is also
independent of the choices of reference frames. On the other hand, 
an object like (\ref{connection}) is both gauge dependent and reference-frame dependent.

All this may give the impression that the gauge transformation (\ref{gauge}) on $A_\mu$
can be equivalently seen as
changes of reference frames. This is indeed so. To see this,
first consider the transformation of matrix elements
(\ref{g1}) under changes of reference frames $R_{AA}$ and $R_{BB}$ for Alice and Bob
effected by the transformation
\begin{equation}\label{Rg}
R_{AB}\mapsto R_{AA}^\dagger R_{AB}R_{BB}.
\end{equation}
We get
\begin{eqnarray}\label{gt1}
\langle \phi_B|V_{BA}R_{AB}|\psi_B\rangle&\mapsto& 
\langle \phi_B|V_{BA}R^\dagger_{AA}R_{AB}R_{BB}|\psi_B\rangle
\nonumber\\
&=&
\langle \phi_{B'}|R_{BB}V_{BA}R^\dagger_{AA}R_{AB}|\psi_{B'}\rangle,
\end{eqnarray}
where we wrote $|\psi_{B'}\rangle=R_{BB}|\psi_B\rangle$
for the transformed state in Bob's frame.
The change of reference frames, as far as the observable (\ref{g1}) is concerned, 
can equivalently 
be viewed as a transformation of $V_{AB}$:
\begin{equation}\label{Vt}
V_{AB}\mapsto R_{AA} V_{AB}R_{BB}^\dagger,
\end{equation}
while leaving $R_{AB}$ invariant.

Next consider the transformation of observables of the form (\ref{g2}):
\begin{eqnarray}\label{gt2}
\langle \phi_A| V_{AB}U_{BB}V_{BA}|\psi_A\rangle\mapsto
\langle \phi_{A'}| R_{AA}V_{AB}R_{BB}^\dagger U'_{BB} R_{BB}V_{BA}R_{AA}^\dagger|\psi_{A'}\rangle,
\end{eqnarray}
where we used the transformation of the unitary $U_{BB}$
\begin{equation}
U_{BB}\mapsto U'_{BB}=R_{BB} U_{BB} R_{BB}^\dagger.
\end{equation}
And again we see that the transformation (\ref{gt2}) can be
effected by
the same transformations (\ref{Vt}) applied to both $V_{AB}$ and $V_{BA}$ instead of transforming $R_{AB}$.
Thus, a change of reference frame $R_{AB}$ can be seen as a gauge transformation of
$A_\mu$ instead.
The easy way to see this is that $V_{AB}$ always occurs in combination
with $R_{BA}$ as $V_{AB}R_{BA}$.

On the other hand, it is clear that gauge-invariance and reference-frame independence are different:
no gauge-dependent quantity is observable, but there are reference-frame dependent observables.
Indeed, those allow observers to learn about other observers' reference frames.
\subsection{Encoding and gauge-invariant measurements}
It is by now well known and appreciated that one can encode information using  
decoherence-free subspaces (see for example \cite{dfs}), 
which in the context of quantum communication \cite{encoding} or quantum computing \cite{dfs} 
protects quantum information
against the decohering effects of joint errors of the generic form $U^{\otimes N}$
acting on $N$ qubits,  where $U\in {\rm SU}(2)$ is an arbitrary and possibly unknown unitary.
We use this technique here to define gauge-invariant measurements (not to be confused with the observables defined in the previous Section) that will be useful in the next Sections.
\subsubsection{Two qubits}
The singlet state
\begin{equation}
|\beta_0\rangle\equiv [|0\rangle|1\rangle-|1\rangle|0\rangle]/\sqrt{2}
\end{equation}
(written in {\em any} basis)
is invariant under $U\otimes U$ for any $U\in {\rm SU}(2)$.
This invariance means nothing more or less than that
the state is reference-frame independent
and gauge-independent.
This leads immediately to the following gauge-invariant POVM on 2 qubits
\begin{equation}\label{POVM}
E_s=|\beta_0\rangle\langle \beta_0|;
\,\,E_t=I^{(2)}-E_s,
\end{equation}
with the subscripts $s$ and $t$ referring to ``singlet'' and ``triplet'',
the usual names for the $J=0$ and $J=1$ angular momentum eigenstates
of 2 spin-1/2 systems.

For later use we
consider the effect on the other three Bell states of $U\otimes U$.
Defining
\begin{eqnarray}\label{Bell}
|\beta_x\rangle&\equiv& [|0\rangle|0\rangle-|1\rangle|1\rangle]/\sqrt{2}\nonumber\\
|\beta_y\rangle&\equiv& [|0\rangle|0\rangle+|1\rangle|1\rangle]/\sqrt{2}\nonumber\\
|\beta_z\rangle&\equiv& [|0\rangle|1\rangle+|1\rangle|0\rangle]/\sqrt{2}\
\end{eqnarray}
we have for any $U\in {\rm SU}(2)$
\begin{eqnarray}\label{UBell}
U\otimes U|\beta_j\rangle= I\otimes U\sigma_j U^\dagger \sigma_j|\beta_j\rangle,
\end{eqnarray}
for $j=0,x,y,z$,
which explains the particular choice of subscripts $x,y,z$ for the various
Bell states in (\ref{Bell}).
It is convenient to consider infinitesimal
transformations 
\begin{eqnarray}
U=\exp(i\epsilon_a\sigma_a)\approx I +i\epsilon_a\sigma_a,	
\end{eqnarray}
acting on the three Bell states $|\beta_x\rangle$,
$|\beta_y\rangle$, and $|\beta_z\rangle$. Using this infinitesimal form it is easy
to see we can then rewrite (\ref{UBell}) as
\begin{eqnarray}
	U\otimes U= \exp(2i\epsilon_a t_a),
\end{eqnarray}
where $t_a$ are the generators for the 3-D ($J=1$) representation of ${\rm SU}(2)$
\begin{eqnarray}
t_x=\left(
\begin{array}{ccc}
	0& 0 &0\\
	0&0& 1\\
	0&1&0
\end{array}
	\right);
\,\,
t_y=\left(
\begin{array}{ccc}
	0& 0 &i\\
	0&0& 0\\
	-i&0&0
\end{array}
	\right);
\,\,t_z=\left(
\begin{array}{ccc}
	0& 1 &0\\
	1&0& 0\\
	0&0&0
\end{array}
	\right),
\end{eqnarray}
written in the basis
$\{|\beta_x\rangle,
|\beta_y\rangle, |\beta_z\rangle\}$.
We can use the same trick, writing the repeated action of a fixed unitary in SU(2)
on any number $N$ of qubits as a direct sum of various different irreducible representations of SU(2),
to find gauge-invariant POVMs for any number of qubits.
\subsubsection{Three qubits}
In the direct-sum representation of SU(2) acting identically on 3 qubits, it is well-known \cite{dfs}
that there are two different $J=1/2$ representations and one $J=3/2$
representation. One 2-D gauge-invariant subspace (labeled by $J=1/2,\lambda=0$)
is spanned by
\begin{eqnarray}
&[	|0\rangle|1\rangle|0\rangle-|1\rangle|0\rangle|0\rangle]/\sqrt{2}\nonumber\\
&[	|1\rangle|0\rangle|1\rangle-|0\rangle|1\rangle|1\rangle]/\sqrt{2},
\end{eqnarray}
where the basis states are defined in such a way that a bit flip on all three qubits on one basis state manifestly
produces the other. 
The other 2-D gauge-invariant subspace (labeled by $J=1/2,\lambda=1$)
is spanned by
\begin{eqnarray}
&[	2|0\rangle|0\rangle|1\rangle-|0\rangle|1\rangle|0\rangle-|1\rangle|0\rangle|0\rangle]/\sqrt{6}\nonumber\\
&[	2|1\rangle|1\rangle|0\rangle-|1\rangle|0\rangle|1\rangle-|0\rangle|1\rangle|1\rangle]/\sqrt{6}
\end{eqnarray}
which can be labeled by ($J=1/2,\lambda=1$).
Finally, the 4-D completely symmetric subspace of three qubits provides the single
$J=3/2$ representation. 

All gauge-invariant POVM on three qubits can then be defined in terms of the projections
onto these three different subspaces
\begin{equation}\label{POVM3}
\{E_{1/2,0},\,E_{1/2,1},\,E_{3/2}\}
\end{equation}
in obvious notation.
\subsubsection{Four qubits}\label{four}
In the basis $\{|\beta_x\rangle,
|\beta_y\rangle, |\beta_z\rangle\}$ defined above we can easily write down 4-qubit states that are invariant
under the joint action of $U^{\otimes 4}$ for any
$U\in {\rm SU}(2)$, i.e., the states with zero angular momentum. Apart from the obviously invariant state 
$|\phi_{0,0}\rangle\equiv |\beta_0\rangle|\beta_0\rangle$
(which we labeled by $J=0,\lambda=0$) we also have the invariant state
\begin{eqnarray}
|\phi_{0,1}\rangle\equiv	\frac{1}{\sqrt{3}}
\left[
|\beta_x\rangle|\beta_x\rangle
-|\beta_y\rangle|\beta_y\rangle+
|\beta_z\rangle|\beta_z\rangle
\right]
\end{eqnarray}
labeled $J=0,\lambda=1$.
Of course, any linear combination of these two states is also invariant. As an example we may rewrite
the states by relabeling the 4 qubits. For instance, if we relabel qubits 1,2,3,4 to 1,3,2,4
we get 
\begin{eqnarray}
	|\phi_{0,0}\rangle_{1234}&=&\left[
	|\beta_0\rangle_{13}|\beta_0\rangle_{24}+
		|\beta_y\rangle_{13}|\beta_y\rangle_{24}-
			|\beta_z\rangle_{13}|\beta_z\rangle_{24}-
				|\beta_x\rangle_{13}|\beta_x\rangle_{24}
	\right]/2
	\nonumber\\
|\phi_{0,1}\rangle_{1234}&=&
\left[
-3|\beta_0\rangle_{13}|\beta_0\rangle_{24}+
		|\beta_y\rangle_{13}|\beta_y\rangle_{24}-
			|\beta_z\rangle_{13}|\beta_z\rangle_{24}-
				|\beta_x\rangle_{13}|\beta_x\rangle_{24}
	\right]/\sqrt{12}\nonumber\\
\end{eqnarray} 
and these two states are obviously invariant as well. 

There are, furthermore, three different 
$J=1$ representations and the one completely symmetric
$J=2$ representation. A specific and detailed exposition of the
corresponding subspaces can be found in, for example, \cite{dfs}.
Here we just label them by their $J$ quantum number and, in the case of $J=1$ by an additional label $\lambda=0,1,2$.
\section{Communication resources}
Just as in Refs.~\cite{harrow,refbit} we can define several different types
of resources to be used in quantum communication protocols.
\subsection{Bipartite entangled states}
We can define an {\em ebit} between two observers $k$ and $l$
as the resource of them sharing a maximally entangled two-qubit state of the form
\begin{eqnarray}\label{singletk}
	I_{kk}\otimes V_{lk}R_{kl}\left[
	|0_k\rangle|1_l\rangle-|1_k\rangle|0_l\rangle\right]/\sqrt{2},	
\end{eqnarray}
where $I_{kk}$ is the identity on $k$.
This {\em ebit} can be created by observer $k$ producing locally the 
singlet state
and subsequently sending it to $l$ over the channel they share.
Similarly we can define an alternative {\em ebit} by reversing the roles
of observers $k$ and $l$, and this leads to the following definition of an {\em ebit}, involving the shared state
\begin{eqnarray}\label{singletl}
V_{kl}R_{lk}\otimes I_{ll}\left[
	|0_k\rangle|1_l\rangle-|1_k\rangle|0_l\rangle\right]/\sqrt{2}.
\end{eqnarray}
An important question is when these two definitions are equivalent, i.e.,
under what conditions these two states can be converted into one another by LOCC (local operations and classical
communication). In general, we can write for any unitary $U$
\begin{eqnarray}
U_{kk}\otimes I_{ll} \left[	|0_k\rangle|1_l\rangle-|1_k\rangle|0_l\rangle\right]/\sqrt{2}
=I_{kk}\otimes U^\dagger_{ll}
\left[|0_k\rangle|1_l\rangle-|1_k\rangle|0_l\rangle\right]/\sqrt{2}	
\end{eqnarray}
In order to convert one {\em ebit} into the other, observer $l$ would have to apply the local
unitary $V_{lk}V_{kl}$. This operation can indeed be implemented by observer $l$ (though observer $k$ would not be able to implement that operation!) by the assumptions of Sections 2 and 3. 
The two definitions of {\em ebit}
are, therefore, equivalent. For the other Bell states (i.e. the triplet states)
the same conclusion holds as they can all be obtained
from the singlet state by locally applying $X$, $Y$, or $Z$: by observer $k$ to the state (\ref{singletk})
and by observer $l$ to the state (\ref{singletl}), respectively. Thus, all Bell states so defined
are equivalent,
just as they are in the usual case where all observers do share reference frames and have perfect quantum communication channels.

The reason that Alice would have to apply the operation $V_{AB}V_{BA}$ while converting an {\em ebit}
she shares with Bob
is obvious: the singlet state can be
transmitted without change from one observer to the next. So, one way to generate an {\em ebit}
is to have Alice create the singlet state and transmit {\em both} qubits
to Bob who then returns one of the qubits to Alice.
The other way to generate an {\em ebit} is for Alice to transmit only one of her qubits to Bob. The difference between the two 
ways of generating an {\em ebit} is 
that one qubit is transmitted from Alice to Bob and back instead of staying in Alice's possession.

In spite of the above observations, we note that not all bipartite entangled states are equivalent.
It is easy to write down maximally entangled states {\em not} equivalent to the 
states considered here, such as the state
	\[
	[|0_k\rangle|1_l\rangle-|1_k\rangle|0_l\rangle]/\sqrt{2},
\]
which observers $k$ and $l$ would be happy to share, as it would allow them to measure
Bell inequalities or to perform teleportation without further ado. Unfortunately for them, the state
cannot be obtained without sharing a reference frame (i.e. knowing $R_{kl}$). 

For yet another type of Bell states {\em not} equivalent to {\em ebits}, 
one can consider Bell states generated by a third observer, and distributed
to two other observers. This example will be examined further below, as it serves as a resource for
quantum data hiding.
\subsection{Tripartite entangled states}
In discussions about three-party entangled states
attention is usually
focused on two different types of states, the GHZ state and the W state. Indeed, these are 
the only 2 inequivalent states that are truly three-party entangled, in the case where all three
parties share a reference frame.
We will see here that without shared reference frames there are infinitely many inequivalent
states with truly three-party entanglement.
The reason is simple enough, there is no gauge-invariant three-qubit state.

Since both the W state and the GHZ state are in the $J=3/2$ representation of SU(2), we really have to consider
only one of these in the present context. We focus on the GHZ state for no reason in particular.
A GHZ state can be prepared from 2 Bell states, shared between three parties.
For instance, if Alice shares one singlet state each with Bob and Charlie, a CNOT applied
by her to her two qubits and measurement of the target qubit in the standard basis
$|0_A\rangle,|1_A\rangle$ leads to a GHZ-like state, namely
\begin{eqnarray}\label{GHZ0}
I_{AA}\otimes V_{BA}R_{AB}\otimes V_{CA}R_{AC} 
\left[	|1_A\rangle|0_B\rangle|0_C\rangle+|0_A\rangle|1_B\rangle|1_C\rangle\right]/\sqrt{2},
\end{eqnarray}
if the result of her measurement is ``0'', and
\begin{eqnarray}\label{GHZ1}
I_{AA}\otimes V_{BA}R_{AB}\otimes V_{CA}R_{AC} 
\left[	|1_A\rangle|0_B\rangle|1_C\rangle+|0_A\rangle|1_B\rangle|0_C\rangle\right]/\sqrt{2},
\end{eqnarray}
when the result is ``1''.

In the former case Alice can simply flip her qubit to convert (\ref{GHZ0}) to the state 
\begin{eqnarray}\label{GHZ}
I_{AA}\otimes V_{BA}R_{AB}\otimes V_{CA}R_{AC} 
\left[	|0_A\rangle|0_B\rangle|0_C\rangle+|1_A\rangle|1_B\rangle|1_C\rangle\right]/\sqrt{2}.	
\end{eqnarray}
Obviously the same state is created
when Alice generates locally a GHZ state
$[|0_A\rangle|0_A\rangle|0_A\rangle+|1_A\rangle|1_A\rangle|1_A\rangle]/\sqrt{2}$
and subsequently sends the second qubit to Bob and the third to Charlie over the channels she shares with them.
In the other case, though, the state (\ref{GHZ1}) generated is genuinely
different and in fact {\em not} equivalent, as either Bob 
or both Alice and Charlie would have to apply the bit flip operation. 
This way they either get
\begin{eqnarray}\label{GHZ1b}
I_{AA}\otimes X_{BB}V_{BA}R_{AB}X_{BB}\otimes V_{CA}R_{AC} 
\left[	|0_A\rangle|0_B\rangle|0_C\rangle+|1_A\rangle|1_B\rangle|1_C\rangle\right]/\sqrt{2}.	
\end{eqnarray}
or
\begin{eqnarray}\label{GHZ1c}
I_{AA}\otimes V_{BA}R_{AB}\otimes X_{CC}V_{CA}R_{AC}X_{CC} 
\left[	|0_A\rangle|0_B\rangle|0_C\rangle+|1_A\rangle|1_B\rangle|1_C\rangle\right]/\sqrt{2}.	
\end{eqnarray}
These two versions of the GHZ state {\em are}
equivalent by construction.

Moreover, we can easily write down four more states corresponding to
(\ref{GHZ})--(\ref{GHZ1c}) but with the roles of the various observers interchanged.
Thus, we have four more GHZ states that are not equivalent.
Generally speaking, all those states are inequivalent because
for almost all unitaries $U^{(1)}$ and $U^{(2)}$ the relation
\begin{eqnarray}
U_{AA}^{(1)}\otimes U_{BB}^{(2)}\otimes I_{CC}  
\left[|0_A\rangle|0_B\rangle|0_C\rangle+|1_A\rangle|1_B\rangle|1_C\rangle\right]/\sqrt{2}\nonumber\\
=
I_{AA}\otimes U^{(1,2)}_{BB,CC}  
\left[	|0_A\rangle|0_B\rangle|0_C\rangle+|1_A\rangle|1_B\rangle|1_C\rangle\right]/\sqrt{2}
\end{eqnarray}
holds with $U^{(1,2)}$ a nonlocal (entangling) unitary operation
on two qubits, that cannot be factorized into a product of two local
unitaries. In fact, the operation $X$ that appears in Eqs.~(\ref{GHZ1b})--(\ref{GHZ1c}) above
is nothing special (other than being related
to the particular protocol used to generate a GHZ state from 3 singlet states),
and when we replace it by an arbitrary unitary not equal to the indentity, 
we get yet another inequivalent state.
Thus there are three continuous sets of inequivalent GHZ states. 
\subsection{Refbits}
Here we generalize the {\em refbit},  a unit of sharing a reference
frame, as defined in \cite{refbit}. 
A useful definition, exploited in the next Section, is as follows:
a {\em refbit} is a single qubit state $|\psi_k\rangle$ in a different observer $l$'s
hands, with $\psi$ chosen ``optimally'' for
the (communication) task at hand. This definition reduces to
that of Ref.~\cite{refbit}, which described the special case where all observers 
agree on the physical meaning of the basis states $|0\rangle$
and $|1\rangle$ but not on the relative phase between them. In that case 
the sharing of an equal superposition of
the basis states is always optimal and constitutes a {\em refbit}. 
\section{Communication protocols}
\subsection{Quantum data hiding}
Suppose Charlie prepares one of two states
$[|0_C\rangle|1_C\rangle\pm |1_C\rangle|0_C\rangle]/\sqrt{2}$,
and distributes this state to Alice and Bob, who consequently end up with a state
\begin{equation}\label{hidden}
V_{AC}R_{CA}\otimes V_{BC}R_{CB}[|0_A\rangle|1_B\rangle\pm |1_A\rangle|0_B\rangle]/\sqrt{2}.	
\end{equation}
Alice and Bob cannot determine which one of the two states
Charlie distributed to them, as they would need to know how Charlie's reference
frame is oriented with respect to theirs. Thus Charlie manages to hide \cite{datahiding} one bit of information (`+' or `-')
in the quantum state (\ref{hidden}). The question is with what additional resources Alice and Bob are able to 
unlock the hidden bit.

First note that Alice and Bob can always unlock the bit if one sends the qubit to the other.
This is a trivial observation if Alice, Bob and Charlie share perfect communication channels, but here
one has to do a little work in order to see this {\em and} it only works for certain
entangles states.
For example, if Bob sends the qubit he received from Charlie to Alice, she ends up with a state
\begin{equation}\label{hidden2}
V_{AC}R_{CA}\otimes V_{AB}V_{BC}R_{CA}[|0_A\rangle|1_A\rangle\pm |1_A\rangle|0_A\rangle]/\sqrt{2}.	
\end{equation}
By applying the unitary (which is indeed measurable by Alice in her reference frame)
\[
U=(V_{AC}V_{CA}) (V_{AB}V_{BC}V_{CA})^\dagger 
\]
to the second qubit, she transforms (\ref{hidden2}) to
\begin{equation}\label{unlocked}
V_{AC}R_{CA}\otimes V_{AC}R_{CA}[|0_A\rangle|1_A\rangle\pm |1_A\rangle|0_A\rangle]/\sqrt{2}.	
\end{equation}
By performing the gauge-invariant two-outcome POVM (\ref{POVM})
she can perfectly distinguish the two states, as one is a triplet state, the other the singlet state.

Note that this trick would not work if Charlie had hidden his bit in the two other Bell states,
since those two states are both triplet states.
This shows the Bell states are not created equal in this context.
Namely, an {\em ebit} created by Charlie is {\em not} equivalent, for Alice and Bob,
to {\em ebits} created by Alice or Bob themselves.

Another resource that would allow Alice and Bob to unlock the bit with some nonzero probability
is two {\em refbits} to be provided by Charlie. For instance, suppose
Charlie sends Alice and Bob a qubit each that he
prepares both in the same state, say $[|0_C\rangle+|1_C\rangle]/\sqrt{2}$. 
We may write the resulting state
in the form
\[
V_{AC}R_{CA}\otimes V_{BC}R_{BA}
\sum_{i,j}C^{\pm}_{ij}|\beta_i\rangle_A\otimes |\beta_j\rangle_B
\] 
in the Bell-state bases $\{|\beta_i\rangle_k\}$ for $i=0,x,y,z$
and $k=A,B$.
One finds then $C^+_{00}=1/\sqrt{8}$ whereas $C_{00}^-=0$.
This implies that 
when both Alice and Bob perform the gauge-invariant POVM
(\ref{POVM}) and both find the result ``singlet'', they have conclusively identified
the '+' bit. A similar conclusion holds when Charlie provides Alice and Bob with orthogonal
{\em refbits}, such as when he transmits
$[|0_C\rangle+|1_C\rangle]/\sqrt{2}$ to Alice and
$[|0_C\rangle-|1_C\rangle]/\sqrt{2}$ to Bob. In that case one finds
$C^-_{00}=1/\sqrt{8}$ whereas $C_{00}^+=0$, so that Alice and Bob can conclusively identify
the '-' bit, when they both find the measurement outcome ``singlet.''
Thus in half of the cases Alice and Bob have a 1/8 chance to unlock the bit,
so that their probability for success is $P_{{\rm success}}=1/16$. But do note that Charlie
can decide to either give them no chance to recover the bit, or a chance of 1/8.
Of course, providing Alice and Bob with infinitely many {\em refbits}
alows them to always unlock the bit.
\subsection{Superdense coding}
Suppose Alice and Bob share an {\em ebit}, a state of the form (\ref{singletk}).
Now Alice performs one of 4 operations $U^{(k)}=I_{AA}, X_{AA}, Y_{AA}, Z_{AA}$
for $k=1\ldots 4$
on her half of the Bell pair, and subsequently sends her qubit to Bob.
He ends up with one of four orthogonal states
\begin{equation}
V_{BA}R_{AB}U^{(k)}_{BB}\otimes V_{BA}R_{AB}\left[
	|0_B\rangle|1_B\rangle-|1_B\rangle|0_B\rangle\right]/\sqrt{2},
\end{equation}
but he can only distinguish the singlet from the
triplet states and nothing more. And so Alice can in fact send not more than 1 bit 
of information with 1 qubit this way. This protocol hardly deserves the name ``superdense coding.''

Alice has to provide at least two {\em refbits}
to allow Bob to receive more bits of information.
The easiest way to analyze this situation is to first consider the case where Alice actually
provides Bob with a maximally entangled state of 2 qubits, say $|\beta_x\rangle$
(note a singlet state $|\beta_0\rangle$ would be a useless resource here), instead of two {\em refbits} 
that are always in a product state.
Bob can then apply, after having applied the two-qubit POVM (\ref{POVM}) 
and having gotten the inconclusive ``triplet'' outcome, 
a gauge-invariant POVM on his 4 qubits.
If Alice in fact applied the $X$ operation, then Bob gets, with probability 1/3, the outcome `$\phi_{0,1}$', which identifies
unambiguously the correct state by excluding the possibilities $Z$ and $Y$.
With other outcomes, Bob learns nothing more about what operation Alice
applied than he already knows from his first measurement.  

Now if Alice sends just two {\em refbits}, say both in the state $|0_A\rangle$,
then Bob receives an equal superposition of $|\beta_x\rangle$ and $|\beta_y\rangle$.
What he then can distinguish unambigously is only the $Z$ vs. the $X$ and $Y$ operations. 
That is, if he gets the `$\phi_{0,1}$' outcome he knows Alice cannot have applied $Z$.
The `$\phi_{0,1}$' outcome occurs with probability 1/6 if Alice in fact applied $X$, and also if in fact she applied
$Y$. So Alice can send more than 1 bit of information by choosing, for example,
$I$ with probability 1/2, $Z$ with probability 1/4, and $X$ or $Y$ with probability
1/4. 
Then in 1/2 of the cases Bob gets only 1 bit, but in 1/24 (=1/6*1/4) of the cases he gets two bits,
which is 1/24th of a bit more than without {\em refbits}. This is not the optimum protocol
but the easiest to explain and an improvement almost worth the name ``superdense coding.''
Of course, in the limit of infinitely many {\em refbits}
the optimum superdense coding protocol allows Alice to send 2 bits to Bob with 1 qubit.
\subsection{No bit commitment}
Consider the following protocol. Alice prepares either the gauge-invariant 4-qubit
state
$|\phi_{0,0}\rangle$ (which we call ``case 0'')
or the gauge-invariant 4-qubit state
$|\phi_{0,1}\rangle$ (``case 1''), both defined in Section \ref{four}.
Then she sends 2 qubits to Bob over the quantum channel they share.
In case 0, she sends the first and the third qubit, in case 1 she either sends
the first and second qubits, or the first and third qubits (without telling him (yet)).
There is only 1 useful
gauge-invariant measurement Bob could perform, namely projecting onto the singlet and
triplet states. From the definitions in Section 4.4.3 we read off that
in case 0, Bob would get the outcome ``singlet'' with probability 1/4.
In case 1, if Alice sends the first and third qubits, he would get that outcome with probability
3/4, whereas if she sends the first and second qubits, he never gets that outcome.
So, if Alice sends the first and third qubits with probability 1/3, Bob
will have a probability of $1/4=1/3\times3/4$
of getting the ``singlet'' outcome, independent of Alice's choice between cases 0 and 1.

Thus the above procedure part can act as
the first stage of a quantum bit commitment protocol, where Alice commits to a bit that Bob cannot 
read. The difference with usual but-commitment schemes is that in the protocol presented above the
choice ``0'' corresponds always to the {\em same} state. For the other choice 
Bob gets, as usual, a mixture of two states. Since that mixture is in fact the same
as the partial density matrix of particles 1 and 3 in the state $|\phi_{0,0}\rangle$,
Alice can in fact rotate one choice to the other, using just one ancilla qubit. Thus she can always cheat perfectly.

This all follows from the general case analyzed in \cite{super}. That Alice needs an ancilla to cheat without going detected, and that in contrast without
ancillas bit commitment would be possible, was shown in \cite{BC}.
\section{Conclusions}
The simple assumption that communication channels are not perfect
leads to additional complications in the theory of reference frames
in quantum communication. This assumption of nonperfect quantum communication channels
arises naturally--- after all how
could observers be sure to share perfect channels when they do not share a reference frame---
but has not been studied so far.

We defined three types of observables, private or public, and reference-frame
dependent or reference-frame independent.
By rewriting certain observables in terms of a gauge vector field,
we showed how reference frame changes can be viewed as gauge transformations. In particular,
``public reference-frame independent observables'' in the quantum communication context
correspond to gauge-independent observables, and one class of those observables in particular corresponds 
to Wilson loops. 

Consideration of quantum communication resources revealed that
not all bipartite maximally entangled states are equivalent to {\em ebits},
and for three-party entangled states the situation is worse: for example,
there are 3 continuous sets of inequivalent GHZ states, that cannot be converted into one another by local operations and classical communication.
Finally, by generalizing the concept of a {\em refbit} \cite{refbit}
one can quantify
how much of a reference frame one has to share in order to be able to
implement superdense coding to some given extent or to unlock a bit in
a quantum data hiding protocol.

\end{document}